\documentclass[aps,prd,preprint,tightenlines,groupedaddress,showpacs]{revtex4}
\usepackage{epsf,epsfig,graphics,graphicx}
\begin{document}

\title{Full-sky correlation functions for CMB experiments with \\
asymmetric window functions}

\author{Kin-Wang Ng}
\email{nkw@phys.sinica.edu.tw}
\affiliation{Institute of Physics \& Institute of Astronomy and Astrophysics,
             Academia Sinica, Taipei, Taiwan 11529, R.O.C.}
\date{\today}

\begin{abstract}
We discuss full-sky convolution of the instrumental beam with the
CMB sky signal in CMB single-dish and interferometry experiments,
using the method~\cite{chall} that
the measured temperature and polarization anisotropies are
defined globally on the group manifold of the three-dimensional rotation
by means of Wigner D-functions. We re-derive the anisotropy and
polarization correlation functions incorporated with asymmetric
window functions, which are then explicitly calculated for
a single-dish elliptical Gaussian beam and an interferometric Gaussian beam.
\end{abstract}

\maketitle

\section{Introduction}

Since the detection of the large-angle temperature anisotropy of the
cosmic microwave background (CMB) by the Cosmic Background Explorer
(COBE) satellite~\cite{smo}, many CMB measurements have reported
detections or upper limits of the CMB temperature anisotropy power spectrum
$C_{Tl}$ over a wide range of $l$~\cite{bou}.
Recently, the Wilkinson Microwave Anisotropy Probe (WMAP)~\cite{ben}
has measured the $C_{Tl}$ in an unprecedented accuracy for $l<800$
and thus made a precise estimation of a number of cosmological parameters.

Polarization of the CMB contains a wealth of information
about the early universe as well.
It can cross-check the measured $C_{Tl}$ and improve
the accuracy in determining the cosmological parameters.
A CMB polarization field can be decomposed into an electric-type $E$ mode
and a magnetic-type $B$ mode~\cite{zalkam}. Recently, the DASI instrument,
a ground-based interferometric array with degree-scale resolution,
has detected the CMB $E$ polarization and $TE$ cross-correlation,
while setting an upper limit on the $B$ polarization~\cite{kov}.
Furthermore, the WMAP has measured the $TE$ power spectrum~\cite{kog},
which is consistent with theoretical predictions based on the measured
CMB anisotropy and indicates a significant large-scale $E$ polarization.
Measuring the CMB polarization has become one of the main goals of CMB
experiments~\cite{tim}. Future ground-based and balloon-borne experiments
and the Planck satellite will unveil detailed features of the CMB anisotropies,
thus allowing one to determine to a high precision the cosmological parameters.

As high-resolution data made by high-sensitivity CMB experiments
comes along, it will be necessary to consider the effect of beam
asymmetry in data analysis in order to make unbiased estimation of
the power spectra and ultimately of the cosmological parameters.
This effect, mainly caused by the off-axis position of the
detector in the focal plane of the telescope, is small and so far
has been largely neglected in CMB data analysis. Recently, the
effect of beam asymmetry has been investigated. Numerical
studies~\cite{bur,wu,chiang,trist} have shown that the typical
difference due to beam asymmetry is of few $\mu K$ in Planck
configuration and increases with the beamwidth and
ellipticity~\cite{bur} and that an azimuthally symmetrized beam
does not bias the power spectrum estimates for most practical
situations~\cite{wu}. In Ref.~\cite{trist}, a fast convolution
algorithm based on the decomposition of the asymmetric beam as a
sum of circular Gaussian functions is developed. On the other
hand, analytic methods have been formulated for treating arbitrary
asymmetric beams in full-sky CMB temperature and polarization
anisotropy experiments by using spin-weighted spherical 
harmonics~\cite{ngliu} and Wigner D-functions~\cite{chall}.
In Ref.~\cite{chall}, the authors have pointed out that for asymmetric 
beam functions the result of convolutions is on the group manifold 
of the three-dimensional rotation and provided an algorithm for 
fast computation of the convolution on this group manifold. 
This method has been applied 
to study the effect of an elliptical Gaussian beam in the estimation
of temperature and polarization correlation functions using
semi-analytic or full numerical integration~\cite{sour} and a
perturbative series in powers of the ellipticity
parameter~\cite{fos}.

In this paper, we will study the effect of beam asymmetry in
full-sky CMB single-dish as well as interferometry experiments. We
will derive the full-sky convolution for a complex asymmetric beam
and define the measured temperature and polarization
anisotropies globally on the group manifold of the
three-dimensional rotation by using Wigner D-functions~\cite{chall}. 
As such, we can extend the standard
anisotropy and polarization correlation functions of domain over
the celestial sphere to those over the rotation group manifold.
The result is then applied to calculate the full-sky correlation
functions for two typical cases: an elliptical Gaussian beam and
an interferometric Gaussian beam. The former is a good
approximation to the actual shape of the window in most of CMB
single-dish experiments and the latter is maximally asymmetric due
to the finite length of the baseline of the interferometer. We
will be able to obtain closed forms for the covariance matrices in
both cases which are directly applicable in the CMB likelihood
data analysis.

The paper is organized as follows. In Sec.~\ref{cmb}, a brief account of the
building block of CMB anisotropies is given.
We present the full-sky convolution of a complex asymmetric beam with the
CMB sky in Sec.~\ref{asymbeam}. The full-sky CMB anisotropy and polarization
correlation functions are re-derived in Sec.~\ref{cf}.
In Sec.~\ref{cases}, we apply the result to study the effect of beam asymmetry
in two cases: a single-dish elliptical Gaussian beam and an
interferometric Gaussian beam. The flat-sky limits of these two cases are
discussed in Sec.~\ref{fsky}. Sec.~\ref{con} is our conclusions.

\section{CMB Temperature and Polarization Anisotropies}
\label{cmb}

Polarized emission is conventionally described in terms of the
four Stokes parameters $(I,Q,U,V)$, where $I$ is the intensity,
$Q$ and $U$ represent the linear polarization, and $V$ describes
the circular polarization. Since circular polarization cannot be
generated by Thomson scattering alone, the parameter $V$ decouples
from the other components and will not be considered. Let us
define $T$ be the temperature fluctuation about the mean; then,
the CMB anisotropies are completely described by $(T,Q,U)$, where
each parameter is a function of the pointing direction $\hat
e(\theta,\phi)$ on the celestial sphere.

Considering the CMB as Gaussian random fields, we can
expand the Stokes parameters as~\cite{zalkam}
\begin{eqnarray}
T(\hat e)&=&\sum_{lm}a_{0,lm}Y_{lm}(\hat e), \nonumber \\
(Q-iU)(\hat e)&=&\sum_{lm}a_{2,lm}\:_{2}Y_{lm}(\hat e), \nonumber \\
(Q+iU)(\hat e)&=&\sum_{lm}a_{-2,lm}\:_{-2}Y_{lm}(\hat e),
\label{expand}
\end{eqnarray}
where $a_{0,lm}$ and $a_{\pm 2,lm}$ are Gaussian random variables, and
$\:_{\pm 2}Y_{lm}$ are spin-2 spherical harmonics given by~\cite{new}~\footnote
{In Ref.~\cite{new}, the sign $(-1)^{s+m}$ is absent.
We have added the sign in order to match the conventional
definition for $Y_{lm}=\:_{0}Y_{lm}$.}
\begin{eqnarray}
\:_{s}Y_{lm}(\theta,\phi)&&=(-1)^{s+m}e^{im\phi}\left[\frac{2l+1}{4\pi}
\frac{(l+m)!}{(l+s)!}\frac{(l-m)!}{(l-s)!}\right]^{\frac{1}{2}}
\sin^{2l} \left(\frac{\theta}{2}\right) \nonumber \\
&&\times\sum_{r}\left(\begin{array}{c}
l-s\\
r\\
\end{array}\right)\left(\begin{array}{c}
l+s\\
r+s-m\\
\end{array}\right)(-1)^{l-s-r}\cot^{2r+s-m}\left(\frac{\theta}{2}\right),
\end{eqnarray}
where
\begin{equation}
\max(0,m-s) \le r\le \min(l-s,l+m).
\end{equation}
Note that $\:_{s}Y_{lm}$ is related to Wigner D-functions
$D^l_{m'm}$~\cite{varsh} by
\begin{equation}
D^l_{m'm}(\psi,\theta,\phi)=e^{-im'\psi} d^l_{m'm}(\theta) e^{-im\phi},
\quad{\rm where}\quad
\:_{s}Y_{lm}(\theta,0)=(-1)^{s+m}\sqrt{\frac{2l+1}{4\pi}} d^l_{-sm}(\theta).
\label{dfct}
\end{equation}
It is straightforward to show the conjugation and symmetry relations
\begin{eqnarray}
\:_{s}Y^*_{lm}(\theta,\phi)&=&(-1)^{s+m}\:_{-s}Y_{l-m}(\theta,\phi),
\label{conj} \\
D^l_{m'm}(\psi,\theta,\phi)&=&(-1)^{m'+m}D^l_{mm'}(\phi,\theta,\psi).
\label{sym}
\end{eqnarray}
Isotropy in the mean guarantees the ensemble averages:
\begin{eqnarray}
\left<a^{*}_{0,l'm'}a_{0,lm}\right>&=&C_{Tl}\delta_{l'l}\delta_{m'm},
\nonumber \\
\left<a^{*}_{2,l'm'}a_{2,lm}\right>&=&(C_{El}+C_{Bl})\delta_{l'l}
\delta_{m'm}, \nonumber \\
\left<a^{*}_{2,l'm'}a_{-2,lm}\right>&=&(C_{El}-C_{Bl})
\delta_{l'l}\delta_{m'm}, \nonumber \\
\left<a^{*}_{0,l'm'}a_{2,lm}\right>&=&-C_{Cl}\delta_{l'l}\delta_{m'm},
\label{enav}
\end{eqnarray}
where $C_{Tl}$, $C_{El}$, $C_{Bl}$, and $C_{Cl}$
are respectively the anisotropy, E polarization, B polarization,
and TE cross correlation angular power spectra.

\begin{figure}
\leavevmode
\hbox{
\epsfxsize=3.0in
\epsffile{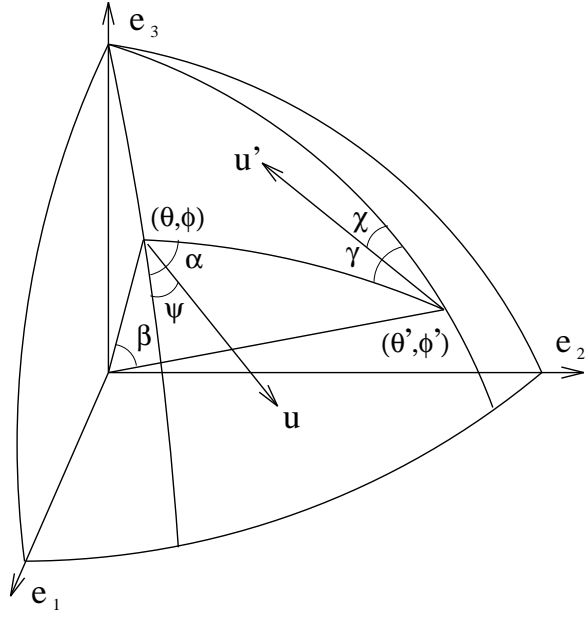}}
\caption{Spherical coordinates showing two unit vectors
$\hat e'(\theta',\phi')$ and $\hat e(\theta,\phi)$ with separation angle
$\beta$. The angles between the great arc connecting the two points and
the longitudes are $\gamma$ and $\alpha$. $\vec u$ is an arbitrary
reference axis on the asymmetric beam pattern or an interferometric
baseline vector tangential to the sphere with the orientation angle $\psi$.}
\label{fig1}
\end{figure}

Consider two pointings $\hat e$ and $\hat e'$ on the celestial sphere.
By making use of Eq.~(\ref{enav}) and
the generalized addition theorem~\cite{ngliu}
\begin{equation}
\sum_{m}\:_{s_1}Y^*_{lm}(\theta',\phi')\:_{s_2}Y_{lm}(\theta,\phi)
        =\sqrt{\frac{2l+1}{4\pi}}(-1)^{s_1-s_2}
           \:_{-s_1}Y_{ls_2}(\beta,\alpha)e^{-is_1\gamma},
\label{addition}
\end{equation}
we find the two-point correlation functions~\footnote
{We have corrected the sign errors in the results found in Ref.~\cite{ngliu},
where $\alpha\to -\alpha$ and $\gamma\to -\gamma$ should be made
in Eqs.~(4.6)-(4.9), (6.3), and (9.2).}
\begin{eqnarray}
&&\left<T^{*}(\hat e')T(\hat e)\right>
  =\sum_l\frac{2l+1}{4\pi}C_{Tl} P_l(\cos\beta), \nonumber \\
&&\left<T^{*}(\hat e')\left[Q(\hat e)+iU(\hat e)\right]\right>
  =-\sum_l\frac{2l+1}{4\pi}\sqrt{\frac{(l-2)!}{(l+2)!}}C_{Cl}
   P^2_l(\cos\beta) e^{-2i\alpha}, \nonumber \\
&&\left<\left[Q(\hat e')+iU(\hat e')\right]^*
  \left[Q(\hat e)+iU(\hat e)\right]\right>
  =\sum_{l}\sqrt{\frac{2l+1}{4\pi}}(C_{El}+C_{Bl})
   \:_{2}Y_{l-2}(\beta,0)e^{-2i(\alpha-\gamma)}, \nonumber \\
&&\left<\left[Q(\hat e')-iU(\hat e')\right]^*
  \left[Q(\hat e)+iU(\hat e)\right]\right>
  =\sum_{l}\sqrt{\frac{2l+1}{4\pi}}(C_{El}-C_{Bl})
   \:_{2}Y_{l2}(\beta,0)e^{-2i(\alpha+\gamma)},
\label{cf1}
\end{eqnarray}
where $\beta$, $\alpha$, and $\gamma$ are the angles defined in Fig.~\ref{fig1}.
Therefore, the statistics of the CMB anisotropy and polarization is fully
described by the four independent power spectra
or their corresponding correlation functions. The details about the evaluation
of the power spectra can be found in Ref.~\cite{zalkam}.

\section{Full-sky Complex Asymmetric Beam Convolution}
\label{asymbeam}

In realistic CMB observations, as a result of the finite beam size of the
antenna and the beam switching mechanism, a measurement is actually a
convolution of the antenna response with the CMB Stokes parameters,
\begin{equation}
X_s^{\rm map}(\hat e,\hat u) = \int d{\hat e'} R(\hat e';\hat e,\hat u)
                             X_s(\hat e'),
\end{equation}
where $X_0=T$, $X_{\pm 2}=Q\mp iU$, and $R(\hat e';\hat e,\hat u)$
denotes the response function with the
pointing direction $\hat e$ and the orientation direction $\hat u$
(${\vec u}=u{\hat u}$, see Fig.~\ref{fig1}). 
The orientation angle is denoted by $\psi$ and therefore we have
$0\le\theta\le\pi$, $0\le\phi<2\pi$, and $0\le\psi<2\pi$. 
Note that $R(\hat e';\hat e,\hat u)$ may be an arbitrary complex function.
Expanding
\begin{equation}
\:_{s}Y_{lm}(\theta',\phi')=\sqrt{\frac{4\pi}{2l+1}}
\sum_{m'}\:_{s}Y_{lm'}(\beta,\alpha)\,e^{is\gamma}\:_{-m'}Y_{lm}(\theta,\phi)
\end{equation}
in Eq.~(\ref{expand}), we obtain the full-sky convolution for an arbitrary
beam function as~\cite{ngliu}
\begin{equation}
X_s^{\rm map}(\theta,\phi,\psi)=\sum_{lmm'}
a_{s,lm}\:_{-m'}Y_{lm}(\theta,\phi)
\sqrt{\frac{4\pi}{2l+1}} \int \sin\beta\,d\beta d\alpha\,
R(\beta,\alpha,\psi)\:_{s}Y_{lm'}(\beta,\alpha)\,e^{is\gamma}.
\label{integral}
\end{equation}
In typical CMB measurements, the field of view is small such that we can make
a {\em local} flat-sky approximation $\alpha=\gamma$. After changing
the integration variable $\alpha\to \alpha-\psi$ in Eq.~(\ref{integral})
and using Eq.~(\ref{sym}), the orientation angle $\psi$ of the response function
is absorbed in the new variable and we thus obtain 
\begin{eqnarray}
&&X_s^{\rm map}(\theta,\phi,\psi)=e^{is\psi}\sum_{lmm'} a_{s,lm}
D^{l*}_{mm'}(\phi,\theta,\psi) \sqrt{\frac{2l+1}{4\pi}}\:_{s}b_{lm'},
\label{main} \\
&&\:_{s}b_{lm'}=\sqrt{\frac{4\pi}{2l+1}} \int \sin\beta\,d\beta d\alpha\,
R(\beta,\alpha)\:_{s}Y_{lm'}(\beta,\alpha)\,e^{is\alpha}.
\nonumber
\end{eqnarray}
This equation was first derived in Ref.~\cite{chall} by using an alternative
way. From the conjugation relation~(\ref{conj}),
we have $\:_{s}b^*_{lm}= (-1)^{s+m}\:_{-s}b_{l-m}$.
We define a spin-$s$ window function
\begin{equation}
\:_{s}W_l\equiv \sum_{m}\left\vert \:_{s}b_{lm} \right\vert^2.
\label{wf}
\end{equation}
For a simple single-dish experiment with an axisymmetric
Gaussian response function given by
\begin{equation}
R^G(\beta,\alpha)=\frac{1}{2\pi\sigma_b^2}
                  \exp\left(-\frac{\beta^2}{2\sigma_b^2}\right),
\label{grf}
\end{equation}
where $\sigma_b\ll 1$ is the Gaussian beamwidth, it can be shown from
Eq.~(\ref{main}) that~\cite{ngliu,chall}
\begin{eqnarray}
&&\:_{s}b^G_{lm}=\:_{s}W_l^{G{1\over2}} \delta_{-m,s}\,, \label{gb} \\
&&X_s^{\rm map}(\theta,\phi)=\sum_{lm} a_{s,lm} \:_{s}W_l^{G{1\over2}}
                           \:_{s}Y_{lm}(\theta,\phi), \label{gX} \\
&&{\rm where}\quad \:_{s}W_l^G =
\exp\left\{-\left[l(l+1)-s^2\right]\sigma_b^2\right\}.
\label{gwf}
\end{eqnarray}
For $s=0,\pm2$ and $\sigma_b\to 0$, the Gaussian window function
$\:_{s}W_l^G\to 1$ and Eq.~(\ref{gX}) reduces to Eq.~(\ref{expand}).

\section{Correlation Functions on Rotation Group Manifold}
\label{cf}

The result in Eq.~(\ref{main}) shows that CMB measurements with
asymmetric beam, although the sky is approximated as flat locally,
can be interconnected globally on the celestial
sphere extended with a circle space of the orientation angle of the
asymmetric beam. More precisely, the observed anisotropies are well-defined
on the group manifold of the three-dimensional rotation~\cite{chall}.
Let us denote a point on the manifold by
$\vec r=(\hat e, \hat u)=(\theta,\phi,\psi)$, then
by making use of Eqs.~(\ref{enav}), (\ref{addition}), and (\ref{main}),
we obtain the two-point correlation functions
(see also Refs.~\cite{sour,fos})
\begin{eqnarray}
\left< X_{s'}^{\rm map}\:^*({\vec r}\,') X_s^{\rm map}(\vec r)\right>
=&&e^{-is'\psi'}e^{is\psi}\sum_{lm'm} \sqrt{\frac{2l+1}{4\pi}} C_{s's,l}
 (-1)^{m'+m} \:_{m'}Y_{l-m}(\beta,0)\,\times \nonumber \\
&& e^{-im(\alpha-\psi)} e^{im'(\gamma-\psi')}
 \:_{s'}b^*_{lm'} \:_{s}b_{lm},
\label{cf2}
\end{eqnarray}
where $C_{00,l}=C_{T,l}$, $C_{0-2,l}=-C_{C,l}$,
and $C_{\mp 2 -2,l}=C_{E,l}\pm C_{B,l}$.
For a Gaussian beam given in Eq.~(\ref{grf}), substituting the result from
Eq.~(\ref{gb}) for both $\:_{s'}b_{lm'}$ and $\:_{s}b_{lm}$ in Eq.~(\ref{cf2}),
we obtain
\begin{equation}
\left<X_{s'}^{\rm map}\:^*({\vec r}\,') X_s^{\rm map}(\vec r)\right>
=\sum_l \sqrt{\frac{2l+1}{4\pi}} C_{s's,l}
 \:_{-s'}Y_{ls}(\beta,0) e^{is\alpha} e^{-is'\gamma}
 \:_{s'}W_l^{G{1\over2}} \:_{s}W_l^{G{1\over2}},
\end{equation}
where $\:_{s}W_l^G$ is the Gaussian window function given in Eq.~(\ref{gwf}).
This reduces to the correlation functions in Eq.~(\ref{cf1})
as the Gaussian beamwidth $\sigma_b\to 0$.

To write Eq.~(\ref{cf2}) in a compact form, we define
${\bar X}_s^{\rm map}(\vec r)=e^{-is\psi} X_s^{\rm map}(\vec r)$,
${\bar\alpha}=\alpha-\psi$, and ${\bar\gamma}=\gamma-\chi$ where
$\psi'=\pi+\chi$. After using Eq.~(\ref{dfct}), we obtain
\begin{equation}
\left<{\bar X}_{s'}^{\rm map}\:^*({\vec r}\,')
{\bar X}_s^{\rm map}(\vec r)\right>
= \sum_{lm'm} \frac{2l+1}{4\pi} C_{s's,l} (-1)^{m'}
D^l_{mm'}({\bar\alpha},\beta,-{\bar\gamma}) \:_{s'}b^*_{lm'} \:_{s}b_{lm}.
\label{cf3}
\end{equation}
Note that ${\bar\alpha}$ and ${\bar\gamma}$ are respectively the angles
between the orientation directions at the
pointings $\hat e$ and $\hat e'$ and the great arc
connecting the two pointing directions (see Fig.~\ref{fig1}).
Furthermore, the root-mean-square anisotropy and polarization fluctuations
are given by the correlation functions~(\ref{cf3}) at zero lag. By taking
${\vec r}\,' \to \vec r$ and making use of the limiting property
(Eq.~(4.16.2) of Ref.~\cite{varsh})
\begin{equation}
D^l_{mm'}(\alpha,0,\gamma) = \delta_{mm'} e^{-im(\alpha+\gamma)},
\end{equation}
we have ${\bar\alpha}-{\bar\gamma}=-\pi$ and
\begin{equation}
\left< \left\vert X_s^{\rm map}(\vec r)\right\vert^2\right>
= \sum_l \frac{2l+1}{4\pi} C_{ss,l} \:_{s}W_l,
\label{rms}
\end{equation}
where $\:_{s}W_l$ is given by Eq.~(\ref{wf}).

Eq.~(\ref{cf3}) is useful for constructing the covariance matrices
in the likelihood analysis of CMB data made by an asymmetric beam.
In the next section, we will show that in most practical cases the
summation in Eq.~(\ref{cf3}) can be largely reduced to a tractable one.
For examples, in the case of a slightly elliptical Gaussian beam,
the summation over $m$ for a fixed $l$ converges very fast
as $|\:_{s}b_{lm}|/|\:_{s}b_{l0}|$ falls off rapidly with increasing $|m|$.
In the case of an interferometric beam with a long baseline
$|\:_{s}b_{lm}|$ is approximately dependent on $l$ only.

\section{Asymmetric Window Functions: Two Cases}
\label{cases}

In deriving the main result~(\ref{main}), we have approximated the sky as
locally flat. This is a good approximation for CMB experiments with high
spatial resolution or small fields of view. We can further evaluate
the integral for $\:_{s}b_{lm}$ in Eq.~(\ref{main})
by use of the approximation (Eq.~(4.18.1.2) of Ref.~\cite{varsh}),
\begin{equation}
d^l_{-sm}(\theta) \simeq (-1)^{s+m} J_{s+m}(l\theta)\quad{\rm for}\quad l\gg 1,
\label{dJ}
\end{equation}
where $J_n(x)$ are Bessel functions and we have used the property
$J_{-n}(x)=(-1)^n J_n(x)$.
Below we discuss two typical cases in CMB experiments.

\subsection{Single-dish Elliptical Gaussian Beam}

For most of simple CMB single-dish experiments, the response function can be
approximated by an elliptical Gaussian beam,
\begin{equation}
R^{EG}(\beta,\alpha)=\frac{1}{2\pi\sigma_x\sigma_y}
\exp\left(-\frac{x^2}{2\sigma_x^2}-\frac{y^2}{2\sigma_y^2}\right),
\label{ebeam}
\end{equation}
where $x=\beta\cos\alpha$, $y=\beta\sin\alpha$, and $\sigma_x$ and $\sigma_y$
are the beamwidths in the major and minor axes respectively. Here we assume
$\sigma_x\ll 1$ and $\sigma_y\ll 1$.
Substituting the beam~(\ref{ebeam}) in Eq.~(\ref{main}), making
the approximation~(\ref{dJ}) in the integral, and using Eqs.~(3.915.2) and
(6.651.6) of Ref.~\cite{grad} to perform the $\alpha$-integration and
$\beta$-integration respectively, we obtain
\begin{eqnarray}
\:_{s}b^{EG}_{lm}\simeq&& {1\over 2}\left[1+(-1)^{s+m}\right]
                     \exp\left[-l^2(\sigma_x^2+\sigma_y^2)/4\right]
               I_{(s+m)/2}\left(l^2(\sigma_x^2-\sigma_y^2)/4\right) \nonumber\\
&&{\rm for}\quad s+m>-1\quad{\rm and}\quad 1>(\sigma_x^2-\sigma_y^2)/\sigma_x^2,
\label{egb}
\end{eqnarray}
where $I_n(x)$ are modified Bessel functions. For $s=0$, the
result coincides with that for an unpolarized elliptical Gaussian beam
found in Ref.~\cite{chall2}.
In the limit of $\sigma_x\to\sigma_y$,
since $I_{(s+m)/2}(0)=\delta_{-m,s}$, we obtain that
$\:_{s}b^{EG}_{lm}\simeq \exp(-l^2\sigma_y^2/2)\delta_{-m,s}$. This is
expected for an Gaussian beam as discussed in Eq.~(\ref{gb}).
From Eq.~(\ref{egb}), the elliptical Gaussian window function is given by
\begin{eqnarray}
\:_{0}W_l^{EG}&\simeq&    \exp\left[-l^2(\sigma_x^2+\sigma_y^2)/2\right]
    \left[I_0^2(z)+2\sum_{n=1}^{[l/2]}I_n^2(z)\right], \\
\:_{\pm2}W_l^{EG}&\simeq& \exp\left[-l^2(\sigma_x^2+\sigma_y^2)/2\right]
    \left[I_0^2(z)+2\sum_{n=1}^{[l/2]-1}I_n^2(z)+I_{[l/2]}^2(z)+
    I_{[l/2]+1}^2(z)\right],
\end{eqnarray}
where $z=l^2(\sigma_x^2-\sigma_y^2)/4$ and $[l/2]$ denotes the integer part
of $l/2$. Making use of the relation (Eq.~(8.537.2) of Ref.~\cite{grad})
\begin{equation}
\sum_{k=-\infty}^{\infty} J_k(z) J_{n-k}(z)=J_n(2z)
\quad{\rm and}\quad I_k(z)=i^{-k} J_k(iz),
\label{zt2z}
\end{equation}
we find that for $z<1$,
\begin{equation}
\:_{s}W_l^{EG}\simeq \exp\left[-l^2(\sigma_x^2+\sigma_y^2)/2\right]\,
I_0\left(l^2(\sigma_x^2-\sigma_y^2)/2\right).
\label{egwf}
\end{equation}

\subsection{Interferometric Gaussian Beam}

Recent advancement in low-noise, broadband, GHz amplifiers
has made interferometry a particularly attractive technique for
detecting CMB anisotropies.  An interferometric array is intrinsically a
high-resolution polarimetric instrument, well suited to observing
small-scale polarized intensity fluctuations while being flexible in the
coverage of a wide range of angular scales, with resolution and
sensitivity determined by the aperture of each element of the array
and the baselines formed by the array elements.
In addition, many systematic problems inherent in single-dish
experiments, such as ground and near field atmospheric pickup, and
spurious polarization signal, can be reduced or avoided in interferometry.
Being ground-based, it is controllable, and it can track the sky
for an extensive period of time, as practiced successfully by the DASI team in
measuring the CMB $E$ polarization~\cite{kov}.
Observational strategies of CMB interferometry experiments such as DASI, CBI,
VSA, and AMiBA can be found in Ref.~\cite{par} and references therein.

In contrast to single-dish experiments which measure or differentiate
the signals in individual dishes, an interferometer measures the
correlation of the signals from different pairs of the array elements.
The correlation output is called the complex visibility. In typical
interferometric measurements, the field of view is small so that
the sky can be treated as flat. Therefore, the complex visibility is
simply the two-dimensional Fourier transform of the intensity fluctuations
on the sky convolved with the primary beam of the interferometer.
The capability of directly sampling the Fourier modes
allows for a simple estimation of the anisotropy power spectrum
from visibility data~\cite{hobson} and an efficient separation of the
$E$ and $B$ polarization power spectra~\cite{par2}.
However, as the sky coverage is increased in mosaicking observations
which combine several contiguous pointings of the telescope,
the curvature of the sky becomes significant even though the sky is
locally flat at each pointing.
Here we apply the results in the previous sections to extend the flat-sky
formalism of the complex visibility
to including the curvature effect of the sky.
For further discussions about large-angular-scale
interferometry in which three-dimensional Fourier transforms are involved,
the reader may refer to Ref.~\cite{ng}.

Let us consider a two-element interferometer and a monochromatic
electromagnetic source. The complex visibility is the
time-averaged correlation of the electric field measured by the
two separated antennae pointing in the same direction to the
sky~\cite{thom}. Usually, each antenna has a pair of feeds which
are sensitive to orthogonal circular or linear polarizations. For
instance, if the dual-polarization feeds measure the right and
left circular polarizations, then the output will be the four
correlations $\langle RR^* \rangle$, $\langle RL^* \rangle$,
$\langle LR^* \rangle$, and $\langle LL^* \rangle$. They can be
related to Stokes parameters $(T,Q,U,V)$. Denoting their
associated visibility functions by $(V_T,V_Q,V_U,V_V)$, we have
\begin{eqnarray}
\langle RR^* \rangle &=& V_T + V_V, \nonumber \\
\langle LL^* \rangle &=& V_T - V_V, \nonumber \\
\langle RL^* \rangle &=& V_Q + iV_U \equiv V_{-2}, \nonumber \\
\langle LR^* \rangle &=& V_Q - iV_U \equiv V_{+2},
\end{eqnarray}
where we have neglected the parallactic angle of the feed with
respect to the sky and the leakage from one polarization channel to the
other polarization channel. The visibility functions are given by
\begin{equation}
V_s(\hat e,\vec u) = \frac{\partial B_\nu}{\partial T}
                     \int d{\hat e'}  A(\hat e';\hat e) X_s(\hat e')
                     e^{2\pi i{\vec u}\cdot {\hat e'}},
\label{vis}
\end{equation}
where $\vec u$ is the separation vector (baseline) of the two antennae
measured in units of the observation wavelength,
$A$ denotes the primary beam with the phase tracking center
pointing along the direction $\hat e$,
and $X_s$ is the CMB sky. In Eq.~(\ref{vis}), $\partial B_\nu/\partial T$
is a conversion factor from the CMB temperature fluctuation to
the brightness fluctuation given by
\begin{equation}
\frac{\partial B_\nu}{\partial T} \simeq 99.27
\frac{x^4 e^x}{(e^x-1)^2} {\rm Jy\;sr^{-1}}\;\mu{\rm K^{-1}},
\quad{\rm where}\quad
x\simeq 1.76 \left(\frac{\nu}{100{\rm GHz}}\right),
\end{equation}
where $B_\nu$ is the Planck function of the photon frequency $\nu$.

Therefore, the response function of the interferometer is a complex function
\begin{equation}
R^I(\hat e';\hat e,\vec u)=A(\hat e';\hat e)
                           e^{2\pi i{\vec u}\cdot{\hat e'}}.
\label{irf}
\end{equation}
Usually, the observation wavelength is much smaller than the size of the
primary beam, dictating a small field of view. As such,
for a single pointing, we can make the flat-sky approximation by decomposing
\begin{equation}
\hat e' = \hat e + \vec \beta ,\quad{\rm with}\quad
{\vec \beta} \cdot \hat e =0,
\quad{\rm and}\quad \left\vert {\vec \beta} \right\vert \ll 1.
\label{fsa}
\end{equation}
Here we assume a Gaussian primary beam given by
\begin{equation}
A(\beta,\alpha)=\exp\left(-\frac{\beta^2}{2\sigma_b^2}\right).
\label{pbeam}
\end{equation}
Hence, by writing ${\vec u}\cdot{\hat e'}={\vec u}\cdot{\vec
\beta}=u\beta\cos\alpha$ in Eq.~(\ref{irf}), the interferometric
response function becomes
\begin{equation}
R^{IG}(\beta,\alpha)=A(\beta,\alpha) e^{i 2\pi u\beta\cos\alpha}.
\label{ibeam}
\end{equation}
Note that in Eq.~(\ref{pbeam}) we have adopted the interferometry convention
that the primary beam does not carry the normalization factor
$1/(2\pi\sigma_b^2)$. Moreover, we have assumed a symmetric primary beam
because a small beam asymmetry introduces only higher-order corrections
which can be neglected as long as the length of the baseline is much
bigger than the size of the dish.

Similar to the previous case, substituting the beam~(\ref{ibeam})
in Eq.~(\ref{main}), making the approximation~(\ref{dJ}) in the integral,
and using Eqs.~(3.915.2) and (6.633.2) of Ref.~\cite{grad} to perform the
$\alpha$-integration and $\beta$-integration respectively, we obtain
\begin{equation}
\:_{s}b^{IG}_{lm}\simeq i^{s+m} 2\pi\sigma_b^2
      \exp\left[-\sigma_b^2(l^2+l_u^2)/2\right] I_{s+m}(ll_u\sigma_b^2),
\label{igb}
\end{equation}
where $l_u=2\pi u$ is the peak location of the primary beam.
When the baseline length is much bigger than the antenna size,
we have the limiting form
\begin{equation}
I_{s+m}(ll_u\sigma_b^2) \simeq
e^{ll_u\sigma_b^2}/\sqrt{2\pi ll_u\sigma_b^2}\quad{\rm for}\quad
ll_u\sigma_b^2 \gg 1.
\end{equation}
Therefore, for long baselines, Eq.~(\ref{igb}) can be approximated as
\begin{equation}
\:_{s}b^{IG}_{lm}\simeq i^{s+m} \sqrt{\frac{2\pi\sigma_b^2}{ll_u}}
      \exp\left[-\sigma_b^2(l-l_u)^2/2\right],
\end{equation}
which is actually a Gaussian beam centered at $l=l_u=2\pi u$.
From Eq.~(\ref{igb}), the interferometric Gaussian window function is
given by
\begin{eqnarray}
\:_{0}W_l^{IG}&\simeq& 4\pi^2\sigma_b^4
              \exp\left[-\sigma_b^2(l^2+l_u^2)\right]
              \left[I_0^2(z)+2\sum_{n=1}^l I_n^2(z)\right], \\
\:_{\pm2}W_l^{IG}&\simeq&  4\pi^2\sigma_b^4
    \exp\left[-\sigma_b^2(l^2+l_u^2)\right]
    \left[I_0^2(z)+2\sum_{n=1}^{l-2}I_n^2(z)+\sum_{n=l-1}^{l+2}I_n^2(z)\right],
\end{eqnarray}
where $z=ll_u\sigma_b^2$. Again, making use of Eq.~(\ref{zt2z}),
we find that for $z<1$,
\begin{equation}
\:_{s}W_l^{IG}\simeq 4\pi^2\sigma_b^4\exp\left[-\sigma_b^2(l^2+l_u^2)\right]\,
I_0(2ll_u\sigma_b^2).
\label{igwf}
\end{equation}

\section{Flat-sky Approximation}
\label{fsky}

We have discussed the full-sky CMB temperature and polarization correlation
functions for asymmetric small-scale beams. In the limit of small sky
coverage, their limiting forms must coincide with the results that are
obtained in the flat-sky approximation.
In this limit, the sky can be treated as flat,
being spanned by a two-dimensional vector ${\bf r}$. Hence, the CMB
anisotropy and polarization fields are given by the Fourier transforms
\begin{eqnarray}
T({\bf r}) &=& \int d{\bf u}
      {\widetilde T}({\bf u}) e^{-2\pi i{\bf u}\cdot {\bf r}}, \nonumber \\
Q({\bf r})\pm iU({\bf r}) &=& \int d{\bf u}
\left[{\widetilde E(\bf u)}\mp i{\widetilde B(\bf u)}\right]
e^{\mp i2\phi_{\bf u}} e^{-2\pi i{\bf u}\cdot {\bf r}},
\label{fexpand}
\end{eqnarray}
where $\phi_{\bf u}$ is the phase in the Fourier space given by
the direction angle of ${\bf u}$ and
\begin{eqnarray}
\langle {\widetilde Y}^*({\bf u}){\widetilde Y}({\bf w}) \rangle &=&
S_{YY}(u)\delta({\bf u}-{\bf w})\quad{\rm where}\quad Y=T,E,B, \nonumber \\
\langle {\widetilde T}^*({\bf u}){\widetilde E}({\bf w}) \rangle &=&
S_{TE}(u)\delta({\bf u}-{\bf w}).
\label{fenav}
\end{eqnarray}
The power spectrum $S(u)$ defined in the ${\bf u}$-plane can be related
to the angular power spectrum $C_l$ defined on the sphere by
$l(l+1)C_l/{2\pi}\simeq 2\pi u^2 S(u)$, with $l\simeq 2\pi u$.

\subsection{Single-dish Elliptical Gaussian Beam}

With an elliptical Gaussian beam with orientation angle $\psi$ given by
\begin{equation}
R^{EG}({\bf r},\psi)=\frac{1}{2\pi\sigma_x\sigma_y}
\exp\left[-\frac{r^2\cos^2(\theta-\psi)}{2\sigma_x^2}
          -\frac{r^2\sin^2(\theta-\psi)}{2\sigma_y^2}\right],
\end{equation}
where ${\bf r}=(r\cos\theta,r\sin\theta)$,
the single-dish measurement gives
\begin{equation}
X_s^{\rm map}({\bf r},\psi) = \int d{\bf r'} R^{EG}({\bf r'}-{\bf r},\psi)
                             X_s({\bf r'}),
\end{equation}
where $X_0=T$, $X_{\pm 2}=Q\mp iU$, and ${\bf r}$ is a pointing position
on the sky. Using the expansion~(\ref{fexpand}) and the Fourier transform
of the response function
\begin{equation}
R^{EG}({\bf r},\psi)=\int d {\bf u} {\widetilde R}^{EG}({\bf u},\psi)
e^{-2\pi i{\bf u}\cdot {\bf r}},
\end{equation}
the measured anisotropies can be written as
\begin{equation}
X_s^{\rm map}({\bf r},\psi) = \int d{\bf w}
{\widetilde R}^{EG}(-{\bf w},\psi) {\widetilde X}_s({\bf w})
e^{is\phi_{\bf w}} e^{-2\pi i{\bf w}\cdot {\bf r}},
\end{equation}
where ${\widetilde X}_0={\widetilde T}$ and
${\widetilde X}_{\pm2}={\widetilde E}\pm i{\widetilde B}$.
Hence, using Eq.~(\ref{fenav}) the two-point correlation function
with different orientation angles at each point is given by
\begin{eqnarray}
&&\left< X_{s'}^{\rm map}\:^*({\bf r'},\psi')
X_s^{\rm map}({\bf r},\psi)\right> \nonumber \\
&=& \int dw\,w S_{s's}(w) \int\phi_{\bf w}\, e^{-i(s'-s)\phi_{\bf w}}
  e^{-2\pi i w |{\bf r'}-{\bf r}|\cos\phi_{\bf w}}
{\widetilde R}^{EG}\:^*(-{\bf w},\psi') {\widetilde R}^{EG}(-{\bf w},\psi),
\end{eqnarray}
where $S_{00}=S_{TT}$, $S_{\pm2\pm2}=S_{EE}+S_{BB}$,
$S_{\pm2\mp2}=S_{EE}-S_{BB}$, $S_{0\pm2}=S_{TE}$,
and $\phi_{\bf w}$ is the angle between ${\bf r}-{\bf r'}$ and ${\bf w}$.
Expanding (see Eq.~(8.511.4) of Ref.~\cite{grad})
\begin{equation}
e^{-2\pi i w |{\bf r'}-{\bf r}|\cos\phi_{\bf w}}
=\sum_{m=-\infty}^{\infty} (-i)^m J_m\left(2\pi w \left\vert{\bf
r'}-{\bf r}\right\vert\right) e^{-im\phi_{\bf w}}
\end{equation}
and using Eq.~(3.915.2) of Ref.~\cite{grad}
to perform the $\phi$-integration,
we finally obtain
\begin{eqnarray}
&&\left< X_{s'}^{\rm map}\:^*({\bf r'},\psi') X_s^{\rm map}({\bf r},\psi)
\right> \nonumber \\
=&&\int dw\,w S_{s's}(w) \exp\left[-2\pi^2 w^2(\sigma_x^2+\sigma_y^2)\right]
\sum_{m=-\infty}^{\infty} \pi \left[(-1)^{s'-s}+(-1)^m\right]
e^{-i(s'-s+m)(\psi'+\psi)/2}\,\times \nonumber \\
&& i^m J_m\left(2\pi w\left\vert{\bf r'}-{\bf r}\right\vert\right)
I_{(s'-s+m)/2}\left(2\pi^2 w^2(\sigma_y^2-\sigma_x^2)\cos(\psi'-\psi)\right).
\end{eqnarray}
At zero lag, the non-vanishing contribution is from $m=0$ since
$J_m(0)=\delta_{m,0}$ and thus
\begin{equation}
\left<\left\vert X_s^{\rm map}({\bf r},\psi)\right\vert^2\right>
=\int dw\,w S_{ss}(w) \exp\left[-2\pi^2 w^2(\sigma_x^2+\sigma_y^2)\right]
 2\pi I_0\left(2\pi^2 w^2(\sigma_y^2-\sigma_x^2)\right),
\end{equation}
which reproduces the result as found in Eq.~(\ref{rms}), with
the window function given by Eq.~(\ref{egwf}) and the replacement $l=2\pi w$.
By taking the limit $\sigma_x\to\sigma_y$ and using
$I_{(s'-s+m)/2}(0)=\delta_{-m,s'-s}$, we obtain the standard two-point
correlation functions with a Gaussian window
\begin{equation}
\left< X_{s'}^{\rm map}\:^*({\bf r'}) X_s^{\rm map}({\bf r})\right>
=\int dw\,w S_{s's}(w) \exp\left(-4\pi^2 w^2\sigma_y^2\right)
2\pi (-i)^{s'-s} J_{s'-s}\left(2\pi w |{\bf r'}-{\bf r}|\right).
\end{equation}

\subsection{Interferometric Gaussian Beam}

In the flat-sky approximation~(\ref{fsa}),
the complex visibility~(\ref{vis}) is reduced to
the two-dimensional Fourier transform of
the Stokes parameter multiplied by the primary beam,
\begin{equation}
   V_s({\bf r},{\bf u}) = \frac{\partial B_\nu}{\partial T}
   \int d{\bf r'} A({\bf r'}-{\bf r}) X_s({\bf r'})
   e^{2\pi i{\bf u}\cdot {\bf r'}},
\end{equation}
where ${\bf u}$ is the two-dimensional projection vector
of the baseline between two dishes in the ${\bf r}$-plane and ${\bf r}$ is a
pointing position on the sky. Therefore, we have
\begin{equation}
 V_s({\bf r},{\bf u}) = \frac{\partial B_\nu}{\partial T}
  e^{2\pi i{\bf u}\cdot {\bf r}} \int d{\bf w}
{\widetilde A}({\bf u}-{\bf w}) {\widetilde X}_s({\bf w})
e^{is\phi_{\bf w}} e^{-2\pi i{\bf w}\cdot {\bf r}}.
\end{equation}
Hence, the two-point correlation function with different baselines
at each point is given by
\begin{eqnarray}
\left< V_{s'}^*({\bf r'},{\bf u'}) V_s({\bf r},{\bf u})\right>
&=& \left(\frac{\partial B_\nu}{\partial T}\right)^2
e^{-2\pi i ({\bf u'}\cdot {\bf r'}-{\bf u}\cdot {\bf r})}
 \int dw\,w S_{s's}(w) \,\times\nonumber \\
&& \int\phi_{\bf w}\, e^{-i(s'-s)\phi_{\bf w}}
  e^{-2\pi i w |{\bf r'}-{\bf r}|\cos\phi_{\bf w}}
  {\widetilde A}^*({\bf u'}-{\bf w}){\widetilde A}({\bf u}-{\bf w}).
\end{eqnarray}
Assuming a Gaussian primary beam~(\ref{pbeam})
and using Eqs.~(3.937.1) and (3.937.2) of Ref.~\cite{grad}
to perform the $\phi$-integration, we obtain for $s'\ge s$
\begin{eqnarray}
\left< V_{s'}^*({\bf r'},{\bf u'}) V_s({\bf r},{\bf u})\right>
&=&\left(\frac{\partial B_\nu}{\partial T}\right)^2
e^{-2\pi i ({\bf u'}\cdot {\bf r'}-{\bf u}\cdot {\bf r})}
\int dw\,w S_{s's}(w) \exp\left[-2\pi^2\sigma_b^2(u'^2+u^2+2w^2)\right]
\,\times \nonumber \\
&& 8\pi^3\sigma_b^4\left[\frac{(p+iq)^2+a^2}{p^2+(a+q)^2}\right]^{(s'-s)/2}
I_{s'-s}\left(\sqrt{(p+ia)^2+q^2}\right),
\end{eqnarray}
where
\begin{eqnarray}
a&=&2\pi w|{\bf r'}-{\bf r}|, \nonumber \\
p&=&4\pi^2\sigma_b^2 w(u'\cos\phi_{\bf u'}+u\cos\phi_{\bf u}), \nonumber \\
q&=&-4\pi^2\sigma_b^2 w(u'\sin\phi_{\bf u'}+u\sin\phi_{\bf u}).
\end{eqnarray}
As $|{\bf r'}-{\bf r}|\to 0$ and we set ${\bf r}={\bf 0}$ without loss of
generality, we obtain the single-pointing visibility covariance matrices
\begin{eqnarray}
\left< V_{s'}^*({\bf u'}) V_s({\bf u})\right>
&=&\left(\frac{\partial B_\nu}{\partial T}\right)^2
\int dw\,w S_{s's}(w) \exp\left[-2\pi^2\sigma_b^2(u'^2+u^2+2w^2)\right]
\,\times \nonumber \\
&&8\pi^3\sigma_b^4 e^{-i(s'-s)\phi_U}
I_{s'-s}\left(4\pi^2\sigma_b^2 w |U|\right),
\end{eqnarray}
where $U=|U|e^{i\phi_U}=u'e^{i\phi_{\bf u'}}+u e^{i\phi_{\bf u}}$,
explicitly given by
\begin{eqnarray}
&&|U|=\left[u'^2+u^2+2u'u\cos(\phi_{\bf u'}-\phi_{\bf u})\right]^{1\over2},
\nonumber \\
&&\phi_U=\tan^{-1}\left[(u'\sin\phi_{\bf u'}+u\sin\phi_{\bf u})/
                        (u'\cos\phi_{\bf u'}+u\cos\phi_{\bf u})\right].
\end{eqnarray}
In a special case when $|{\bf u'}|=|{\bf u}|=u$, we have
$|U|=2u\left\vert\cos\left[(\phi_{\bf u'}-\phi_{\bf u})/2\right]\right\vert$
and $\phi_U=(\phi_{\bf u'}+\phi_{\bf u})/2$.
Finally, the root-mean-square visibility is given by
\begin{equation}
\left<\left\vert V_s({\bf r},{\bf u})\right\vert^2\right>
=\left(\frac{\partial B_\nu}{\partial T}\right)^2 8\pi^3\sigma_b^4
 \int dw\,w S_{ss}(w) \exp\left[-4\pi^2\sigma_b^2(u^2+w^2)\right]
 I_0(8\pi^2\sigma_b^2 uw),
\end{equation}
which reproduces the result as found in Eq.~(\ref{rms}), with
the window function given by Eq.~(\ref{igwf}) and the replacement $l=2\pi w$.

\section{Conclusions}
\label{con}

Next-generation CMB experiments with high resolution and
sensitivity will need to take into account the effect of beam
asymmetry in interpreting observational data, while the pipeline
for data analysis in CMB interferometric measurements using the
flat-sky approximation will have to be modified to including the
effect of the sky curvature to deal with future observational data
with large sky coverage. We have thus studied full-sky CMB
anisotropy and polarization measurements with asymmetric beams. 
The measured Stokes parameters are well defined globally on the group
manifold of the three-dimensional rotation and we have derived their
correlation functions on the group manifold. The correlation functions 
are useful for constructing the full covariance matrices in the 
maximum-likelihood data analysis in large-sky CMB experiments with asymmetric 
window functions, especially when the beam is 
highly asymmetric and it is difficult to do beam symmetrization.
Moreover, the domain of the CMB observable field is extended 
from the celestial sphere on which the field is
expanded in terms of spherical harmonics to the rotation group
manifold on which the field is expanded in terms of Wigner
D-functions. As such, unbiased CMB angular anisotropy and polarization power 
spectra can be directly deconvoluted from observational data by converting 
the Wigner D-functions on the rotation group manifold. Work along this line
is in progress.

\begin{acknowledgments}
The author would like to thank the Berkeley Cosmology Group for their useful
discussions and hospitality during his visit,
where part of the work has been done.
This work was supported in part by the National Science Council, ROC
under the Grant NSC93-2112-M-001-013.
\end{acknowledgments}


\begin{references}

\bibitem{smo}
G. F. Smoot {\it et al.}, Astrophys. J. {\bf 369}, L1 (1992).

\bibitem{bou}
For reviews, see F. R. Bouchet as well as A. H. Jaffe,
Proceedings of the 2002 International Symposium in Cosmology and
Particle Astrophysics, Taiwan, edited by X.-G. He {\it et al.}
(World Scientific, Singapore, 2003).

\bibitem{ben}
C. L. Bennett {\it et al.}, Astrophys. J. Suppl. {\bf 148}, 1 (2003).

\bibitem{zalkam}
M. Zaldarriaga and U. Seljak, Phys. Rev. D {\bf 55}, 1830 (1997);
M. Kamionkowski, A. Kosowsky, and A. Stebbins, {\it ibid.} {\bf 55},
7368 (1997).

\bibitem{kov}
J. M. Kovac {\it et al.}, Nature, {\bf 420}, 772 (2002).

\bibitem{kog}
A. Kogut {\it et al.}, Astrophys. J. Suppl. {\bf 148}, 161 (2003).

\bibitem{tim}
P. T. Timbie, J. O. Gundersen, and B. G. Keating, in {\em AMiBA 2001:
High-z Clusters, Missing Baryons, and CMB Polarization},
edited by L.-W. Chen {\it et al.}, ASP Conference Series Vol. 257
(Astronomical Society of the Pacific, San Francisco, CA, 2002);
J. E. Carlstrom {\it et al.}, Proceedings of "The Cosmic Microwave Background
and its Polarization", New Astronomy Reviews, edited by
S. Hanany and K. A. Olive (Elsevier), astro-ph/0308478.

\bibitem{bur}
C. Burigana {\it et al.}, Astron. Astrophys. Suppl. {\bf 130}, 551 (1998).

\bibitem{wu}
J. H. P. Wu {\it et al.}, Astrophys. J. Suppl. {\bf 132}, 1 (2001).

\bibitem{chiang}
L.-Y. Chiang {\it et al.}, Astron. Astrophys. {\bf 392}, 369
(2002); R. Vio {\it et al.}, astro-ph/0309145.

\bibitem{trist}
M. Tristram {\it et al.}, Phys. Rev. D {\bf 69}, 123008 (2004).

\bibitem{ngliu}
K.-W. Ng and G.-C. Liu, Int. J. Mod. Phys. D {\bf 8}, 61 (1999).

\bibitem{chall}
A. Challinor {\it et al.}, Phys. Rev. D {\bf 62}, 123002 (2000);
B. D. Wandelt and K. M. G{\'o}rski, Phys. Rev. D {\bf 63}, 123002 (2001).

\bibitem{sour}
T. Souradeep and B. Ratra, Astrophys. J. {\bf 560}, 28 (2001).

\bibitem{fos}
P. Fosalba, O. Dor{\' e}, and F. R. Bouchet,
Phys. Rev. D {\bf 65}, 063003 (2002).

\bibitem{new}
E. Newman and R. Penrose, J. Math. Phys. {\bf 7}, 863 (1966);
J. N. Goldberg {\it et al.}, {\it ibid.} {\bf 8}, 2155 (1967).

\bibitem{varsh}
See, for example, D. A. Varshalovich, A. N. Moskalev, and V. K. Khersonskii,
{\em Quantum Theory of Angular Momentum} (World Scientific, Singapore, 1988).

\bibitem{grad}
I. S. Gradshteyn and I. M. Ryzhik, {\em Table of Integrals, Series, and Products}
(Academic Press, San Diego, 2000).

\bibitem{chall2}
A. D. Challinor {\it et al.},
Mon. Not. R. Astron. Soc. {\bf 331}, 994 (2002).

\bibitem{par}
C.-G. Park {\it et al.}, Astrophys. J. {\bf 589}, 67 (2003).

\bibitem{hobson}
M. P. Hobson, A. N. Lasenby, and M. Jones,
Mon. Not. R. Astron. Soc. {\bf 275}, 863 (1995).

\bibitem{par2}
C.-G. Park and K.-W. Ng, Astrophys. J. {\bf 609}, 15 (2004).

\bibitem{ng}
K.-W. Ng, Phys. Rev. D {\bf 63}, 123001 (2001).

\bibitem{thom}
A. R. Thompson, J. M. Moran, and G. W. Swenson,
{\em Interferometry and Synthesis in Radio Astronomy}
(Wiley, New York, 1986).

\end{references}
\end{document}